\documentclass{epl}
\usepackage{amssymb}

\usepackage{graphicx}
\usepackage{dcolumn}
\usepackage{bm}

\title{On the magnetic nature of quantum point contacts}
\author{P. S. Cornaglia \and C. A. Balseiro}
\shortauthor{P. S. Cornaglia \etal}
\institute{Instituto Balseiro and Centro At\'omico Bariloche, Comisi\'on
Nacional de Energ\'{\i}a At\'omica, 8400 San Carlos de Bariloche, R\'{\i}o
Negro, Argentina.\\ 
}

\begin{document}
\maketitle

\begin{abstract}
We present results for a model that describes a quantum point contact. We
show how electron-electron correlations, within the unrestricted
Hartree-Fock approximation, generate a magnetic moment in the point contact.
Having characterized the magnetic structure of the contact, we map the
problem onto a simple one-channel model and calculate the temperature
dependence of the conductance for different gate voltages. Our results are
in good agreement with experimental results obtained in GaAs devices and
support the idea of Kondo effect in these systems.
\end{abstract}

Ballistic transport in quantum wires with ideal contacts leads naturally to
conductance quantization in units of $2e^{2}/h$~\cite{wees1988, cuant2, data}%
. This notable result has become one of the paradigms of mesoscopic physics.
By a quantum wire we understand conductor that is so thin that the
quantization of the transverse modes becomes important. In short quantum
wires we may also expect some quantization effects along the wire axis 
\cite{aust2}. In some of these short wires and in quantum point contacts, as
the gate voltage is varied, together with the $2e^{2}/h$ plateaus the
conductance shows some structure near $0.7(2e^{2}/h)$ \cite
{cerosiete,cerosiete1,crone,aust1}. This structure has been systematically observed
GaAs devices and is known as the $0.7$ anomaly. Recent experiments presented
evidence on the magnetic nature of this structure: for gate voltages at the $%
0.7$ anomaly the conductance increases as the temperature is lowered, an
effect that may be interpreted as a signature of Kondo effect \cite{crone}.
The Kondo picture is supported by the scaling of the experimental data, that
close to the 0.7 anomaly shows a universal temperature dependence. Moreover
the low temperature differential conductance presents a narrow structure at
zero bias voltage, which splits with an in-plane magnetic field as due to a
Kondo resonance. The conventional Kondo effect is the magnetic screening of
a localized moment \cite{Hewson}, if this effect is taking place in the
quantum point contact (QPC), then the main question is: where and how is the
magnetic moment generated? Recent spin dependent functional density
calculations suggest the formation of a magnetic moment localized at the QPC
for a device built on GaAs \cite{Berggren, meir1,meir2}. This scenario in
which a magnetic moment is spontaneously generated by the electron-electron
correlations is appealing since it naturally leads to the Kondo physics
observed experimentally, however there are still many open questions related
to the magnetic nature of the QPC and its relation to the 0.7 anomaly.

\begin{figure}[tbp]
\begin{center}
\includegraphics[width=8.0cm,clip=true]{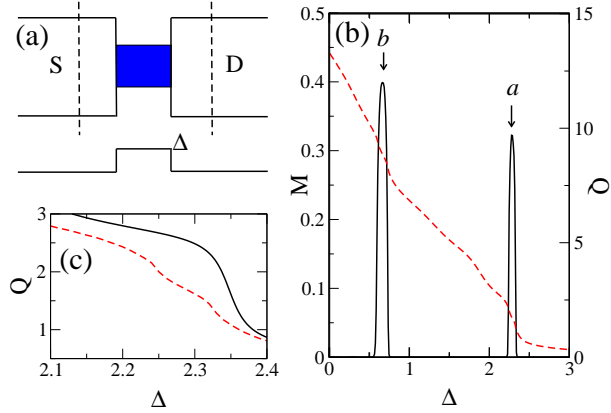}
\end{center}
\caption{(a) Schematic picture of the model for the QPC, the potential at
the constriction is $\Delta$. (b) Magnetic moment (continuous line) and
total charge (dashed line) at the constriction of a $3$-site-wide and $9$%
-site-long QPC. The first conduction channel opens at $a$ and the second at $%
b$, the large magnetization at these points signals the formation of a local
magnetic moment. The local interaction is $U=2.0$ and the Fermi energy is $%
\epsilon_F=-0.95$, with all parameters in units of $t=1$. The source and
drain slabs have a width of $10$ sites. (c) Detail of the total charge
around $a$ (dashed line) compared with a non-interacting system (continuous
line).}
\label{fig1}
\end{figure}

For the sake of simplicity we resort to a one band Hubbard model with local
interactions. 
\begin{equation}
H=\sum_{i,\sigma }\varepsilon _{i}c_{i\sigma }^{\dagger }c_{i\sigma
}+U\sum_{i}c_{i\uparrow }^{\dagger }c_{i\uparrow }c_{i\downarrow }^{\dagger
}c_{i\downarrow }-t\sum_{\langle i,j\rangle ,\sigma }c_{j\sigma }^{\dagger
}c_{i\sigma }  \label{ham}
\end{equation}
here $c_{i\sigma }$ annihilates an electron with spin $\sigma $ at site $i$, 
$U$ is the on-site Coulomb repulsion and the last term in Eq. (\ref{ham})
describes the nearest neighbor hopping with matrix element $t$. The on-site
energy $\varepsilon _{i}$ describes the potential at the QPC. For
simplicity, in what follows we use a square potential parametrized with a
single variable $\Delta $: using the geometry of Fig.~\ref{fig1}(a) we take $%
\varepsilon _{i}=0$ for $i$ in the source and drain, and $\varepsilon
_{i}=\Delta ,\infty $ for $i$ in the contact and at the sides of the contact
respectively. The energy $\Delta $ can be controlled by a gate potential $%
V_{g}$ and we may consider $\Delta \propto -|e|V_{g}$. The square potential
maximizes resonances for states with wavelengths commensurate with the point
contact, nevertheless smoother potentials give similar results \cite{Cornaglia2003}. We look for the ground state of Hamiltonian (1) in the unrestricted
Hartree-Fock approximation for slabs with a QPC as in Fig.~\ref{fig1}(a). In
the central region [between dashed lines in Fig.~\ref{fig1}(a)], the
expectation value of the electron number for each spin and at each site is
calculated self-consistently. Away from the constriction, in the source and
drain, we use the Hartree-Fock solution of a uniform system with
approximately $0.5$ electrons per site. For the results presented below, we
checked that the boundary between the two regions is far enough from the
contact so that the solution becomes independent of its position. Given the
geometry, i.e., the length and the width of the point contact, and the value
of $U$ within a wide interval, we find that for some values of $\Delta$ a magnetic moment is formed in the region of the
constriction. We define the total charge $Q$ and the magnetization $M$ of
the contact as the expectation value of the operators $\hat{Q}%
=\sum_{i,\sigma }^{^{\prime }}c_{i\sigma }^{\dagger }c_{i\sigma }$ and $\hat{%
M}=\frac{1}{2}\sum_{i}^{^{\prime }}\left( c_{i\uparrow }^{\dagger
}c_{i\uparrow }-c_{i\downarrow }^{\dagger }c_{i\downarrow }\right) $ where
the sum is over all the sites in the neck of the QPC [shaded region in Fig. 
\ref{fig1}(a)]. The results are summarized in Fig.~\ref{fig1}(b): for large
negative gate voltages the total charge $Q$ in the point contact is
exponentially small. As $\Delta $ decreases $Q$
increases and for some values of $\Delta$ there is an abrupt
increase in the total charge. At the step-like structures obtained around
the point marked as $a$ in Fig.~\ref{fig1}(b) [see also detail in~\ref{fig1}%
(c)], approximately one electron is transferred from the source and drain to
the point contact, a behavior characteristic of Coulomb blockade. 
Between the two first steps in $Q$ there is a spin 1/2 localized
at the point contact as indicated by the magnetization curve shown also in
Fig.~\ref{fig1}(b). 
In Fig.~\ref{fig1}(c) we present a detail of the behavior of the charge in the QPC as a function of the gate voltage near the point $a$ of Fig.~\ref{fig1}(b) for two values of the local interaction. For the $U=2$ case the entrance of an electron in the QPC is blocked, between the two steps, by the Coulomb interaction. If $U$ is decreased the separation between the steps diminishes linearly, and also the magnetization decreases as the number of electrons in the QPC is less well defined. For $U\lesssim 0.85$ the two steps merge into a $\delta Q\sim 2$ step and the magnetization vanishes.

\begin{figure}[tbp]
\begin{center}
\includegraphics[width=8.0cm,clip=true]{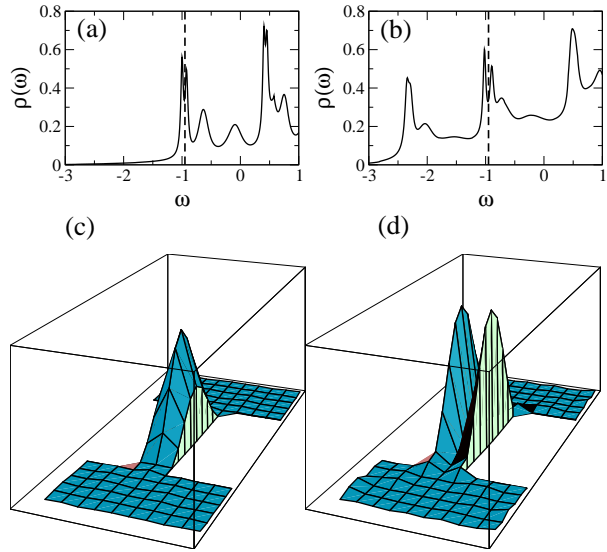}
\end{center}
\caption{(a) Average in the QPC region of the total local density of states
(spin up plus spin down) for the system of Fig.~\ref{fig1}(b) and $%
\Delta=2.27$. (b) Same as (a) with $\Delta=0.66$. The vertical dashed line
in (a) and (b) marks the position of the Fermi level. (c) and (d) spatial
distribution of the magnetization for the (a) and (b) situations
respectively.}
\label{fig2}
\end{figure}

At the point $a$ of Fig.~\ref{fig1}(b), where a spin 1/2
local moment is stable, a narrow resonance is crossing the Fermi energy.
This resonance is clearly seen in the local density of states (LDOS) of the
QPC as shown in Fig.~\ref{fig2}(a). This first resonance corresponds to a
state with no nodes in the QPC as can be inferred from the magnetization of
Fig.~\ref{fig2}(c). Note that the magnetic moment is localized at the QPC.
At the point marked as $b$ in Fig.~\ref{fig1}(b) again a spin 1/2 is
localized in the QPC, this is due to a resonance of the second channel [see Fig.~\ref{fig2}(b)] that
generates a magnetization profile with a node along the mayor axis of the
QPC as shown in Fig.~\ref{fig2}(d). When the chemical potential coincides
with a narrow resonance, the system Hamiltonian can be approximately mapped into that of
an Anderson impurity~\cite{Anderson} with effective interaction $U_{eff}$
and resonance width $\Gamma _{eff}$. If $U_{eff}\gtrsim \pi \Gamma _{eff}$,
in analogy with Anderson criterion for magnetic impurities \cite{Hewson}, we
expect a magnetic moment in the QPC. The effective parameters scale with the
size of the QPC as: $U_{eff}\sim U/A$ where A is the area of the QPC and,
for the geometry of Fig.~\ref{fig1}(a), $\Gamma _{eff}\propto 1/L^{3}$,
where $L$ is the length of the QPC \cite{Cornaglia2003}.

These results deserve a few comments:

\textit{i.- }The behavior of $Q$ as function of $\Delta $ and the occurrence
of a local magnetic moment in the QPC is reproduced for a wide range of
values of the Coulomb repulsion $U$. For the electron densities used up to
now ($\sim 0.5$ electrons per site), the critical value of $U$ that
generates a global instability $U_{c}$ is larger than $8t$. The behavior
presented in Fig.~\ref{fig1}(b) corresponds to $U=2t$, far away from the
global instability. We stress that the model aims to give a qualitative
description with the minimum number of parameters.


\textit{ii.-}The formation of a local magnetic moment is not a peculiarity
of the model used to simulate the potential due to the gate voltage. We have
analyzed different potentials to show that the local instabilities shown in
Fig.~\ref{fig1}(b), are robust for potentials that have a flat region and
produce resonances. Conversely, for wedge-like point contacts~\cite{wees1988}%
, the Hartree-Fock solution does not show a magnetic moment for moderate
values of $U$.

\textit{iii.- }For the geometries of Fig.~\ref{fig1}(a), narrow resonances
are obtained at the bottom of each channel band. Localized magnetic moments
are then obtained each time a new channel becomes active in agreement with
Ref.~\cite{aust1}. The details of the magnitude and spatial
distribution of the magnetic moment depend on the shape of the QPC potential.

\textit{iv.- }The Hartree-Fock approximation, as well as other methods like
the spin dependent functional density approximation~\cite{meir1}, can not be
fully satisfactory since they break the local symmetry. In the exact
solution of the problem, spin fluctuations should recover local rotational
invariance. However as for the original problem of magnetic moment formation
in impurities~\cite{Hewson}, the Hartree-Fock results do indicate the region
in parameter space where we can expect magnetic fluctuations to play a
central role in the low energy physics. A detailed analysis of the local
magnetic instability for different QPC potentials will be
presented elsewhere~\cite{Cornaglia2003}.

\begin{figure}[tbp]
\begin{center}
\includegraphics[width=8.0cm,clip=true]{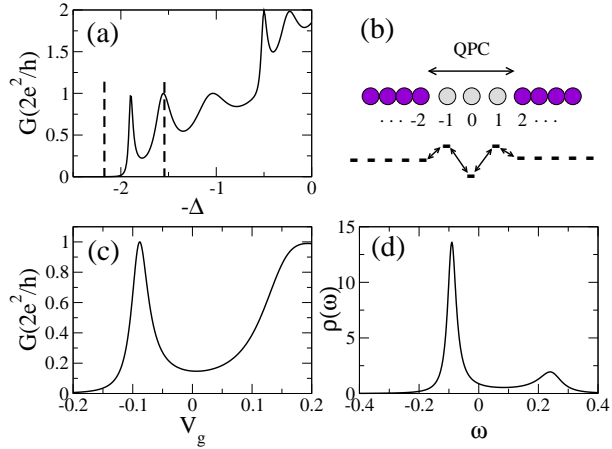}
\end{center}
\caption{(a) Conductance for a $3\times 9$ QPC in a non-interacting system
as a function of the barrier height in the QPC region. (b) Effective 1D
model to describe the first step in (a). (c) Conductance and (d) local
density of states at the central site of (b) for a QPC with $\epsilon_0=-V_g$%
, $\epsilon_1=\epsilon_{-1}=-V_g+0.15$, $V=0.15/\protect\sqrt{2}$, and $%
t_{1,2} = t_{-1,-2} = 0.25$.}
\label{fig3}
\end{figure}

To go beyond the Hartree-Fock approximation and to make a connection with
the Kondo problem a simplification of the model is required. To do so, we
first revisit the transport properties of a non-interacting QPC that can be 
calculated exactly using standard techniques \cite{data,pasw}. The
conductance as a function of the gate voltage $V_{g}$ is shown in Fig.~\ref
{fig3}(a). The results presented in the figure are characteristic of potentials having
a flat region along the mayor QPC axis, the details however depend on the specific potential used to describe de QPC. The conductance for longer QPC
oscillates more rapidly since resonances are narrower and closer in energy.
We note that here, when the conductance in units of $2e^{2}/h$ is smaller
than one, the transmission is due to a single channel. In this one channel
regime, and at low enough temperatures, we can map the problem onto an
effective one-dimensional (1D) model. 

In what follows we present a model that, in the gate voltage interval
indicated with vertical dashed lines in Fig.~\ref{fig3}(a), reproduces the
structure of the conductance and of the density of states. In Fig.~\ref{fig3}(b) a schematic picture of a linear chain with a $3$-site central region
representing the QPC is shown. The Hamiltonian is given by 
\begin{equation}
\mathcal{H}_{1D}^0=\sum_{i=-\infty,\sigma}^\infty \varepsilon_i c_{i\sigma}^\dagger c_{i\sigma} - \sum_{i=-\infty,\sigma}^\infty t_{i,i+1} (c_{i\sigma}^\dagger c_{i+1\sigma}
+ c_{i+1\sigma}^\dagger c_{i\sigma}).
\end{equation}
To reproduce the structure we adjust the hybridization of the central 
site $V=t_{0,1}=t_{-1,0}$, those of the adjacent sites $%
t_{1,2}=t_{-2,-1}$ and the diagonal energies $\epsilon_0$ and $%
\epsilon_{-1}=\epsilon_1$. Only the diagonal energies in the 3-site central region are modified by the gate potential (they are shifted by $V_{g}$). The other paremeters are: $\epsilon_i=0$ for $|i|>1$ and $t_{i,i+1}=1$ for $i\neq\{-2,-1,0,1\}$.
The conductance as a function of a gate voltage for this 1D
model, shown in Fig.~\ref{fig3}(c), reproduces the first structures of Fig.~%
\ref{fig3}(a). The first peak in the conductance is due to a resonant state
crossing the Fermi level. This resonance is observed in the local density of
states (LDOS) of the central site as shown in Fig.~\ref{fig3}(d). Having
this non-interacting 1D model that mimics the transport properties of the
QPC, we can now turn on the interaction that generates a local moment. The
local moment should be associated with the first resonance in the LDOS and
then we include a local interaction at the central site: $\mathcal{H}_{1D}=\mathcal{H}_{1D}^0 + Uc_{0\uparrow}^\dagger c_{0\uparrow}c_{0\downarrow}^\dagger c_{0\downarrow}$.
The temperature dependence of the conductance for this model in the linear regime can be calculated exactly using the numerical renormalization group~\cite{NRG,NRG1}. We
proceed as in Ref.~\cite{nos} to put the conductance in terms of the
spectral density of the correlated site $\rho _{0}(\omega )$.

\begin{equation}
G(T)=\frac{2e^{2}}{h}2 \pi \int d\omega (-\frac{\partial f(\omega )}{%
\partial \omega })\Gamma (\omega )\rho _{0}(\omega )
\end{equation}
where $f(\omega )$ is the Fermi function, and $\Gamma(\omega )=\pi V^{2}\rho
_{1}(\omega )$ with $\rho_{1}(\omega )$ the spectral density of site 1 with $%
V=0$.

\begin{figure}[tbp]
\begin{center}
\includegraphics[width=8.0cm,clip=true]{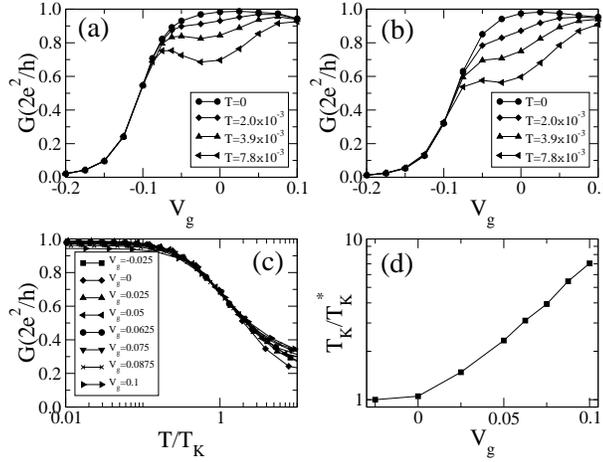}
\end{center}
\caption{(a) Conductance as a function of the gate voltage $V_g$ for the
one-dimensional model of Fig. 2(b) with $U=1.0$ in the central site and the
other parameters as in Fig.~\ref{fig2}(c). (b) Same as (a) with $V=0.2/%
\protect\sqrt{2}$ and $\epsilon_1=\epsilon_{-1}=-V_g+0.25$. (c) Conductance
as a function of the scaled temperature for the parameters in (a). (d) Kondo
temperature obtained from the scaling in (c), $T_K^*=T_K(V_g=-0.025)=1.35
\times 10^{-3}$.}
\label{fig4}
\end{figure}

The results are shown in Fig.~\ref{fig4}(a) and~\ref{fig4}(b) for two sets of parameters that correspond to a long QPC (sharp resonances) and a short QPC (wide resonances) respectively. At intermediate temperatures the conductance presents a plateau at about $0.7(2e^{2}/h)$.
This structure is not universal, depending on the parameters a more
pronounced maximum can be obtained and the whole structure may be slightly
shifted up or down. As the temperature decreases the conductance increases toward $2e^{2}/h$, the plateau shifts away from $0.7(2e^{2}/h)$ and disappears. These numerical results are notably similar to the experiments of Ref.~\cite{crone} and also agree with the experiments of Ref.~\cite{aust2} where sharper structures are found for longer QPC and a maximum, instead of a plateau, is obtained in the conductance.

When the first resonance of the QPC is above the Fermi level [$V_g \lesssim -0.12$ for the parameters of Fig. \ref{fig4}(a)] the conductance and the charge in the QPC are small. An increase in $V_g$ pushes the levels of the QPC below the Fermi level, favoring the charging of the QPC. At $V_g\sim -0.08$ the Fermi level of the leads coincides with the first resonance in the density of states of the QPC. The charge in the QPC is $\sim 1$, and it has a spin $1/2$.
For $V_g \gtrsim -0.07$ the magnetic moment is formed and at low temperatures there is an increase in the conductance as the Kondo effect develops.
 The conductance as a function of temperature for
different values of the gate voltage is shown in Fig.~\ref{fig4}(c). At low
temperatures, all the curves collapse into a single one when the temperature
is properly scaled. We define the Kondo temperature $T_{K}$ as the scaling
parameter. Contrary to what is obtained in the conventional Anderson model~%
\cite{Hewson}, in our mapping the Kondo temperature increases as the gate
voltage increases and the resonance is pushed \textit{away} from the Fermi
level [see Fig.~\ref{fig4}(d)]. This is in agreement with the experimental observations and is due to
an increase of the density of states at the Fermi energy as $V_{g}$
increases and the second longitudinal resonance approaches the Fermi level.
The main difference with the conventional Anderson model is the dependence of the hybridization function $\Gamma(\omega)$ on $V_g$. In the Anderson model $\Gamma(\omega)$ is usually taken as a constant $\Gamma(\omega)=\Gamma(\epsilon_F)$ to describe e.g. quantum dots. In this case $\Gamma(\epsilon_F)$ increases as the gate voltage increases, leading to an enhancement of the Kondo temperature.
We presented results for $U=1.0$, however similar results are obtained for $U\gtrsim 0.25$. To be able to observe the Kondo behavior the local interaction $U$ must be larger than the minimum value of the hybridization $\Gamma(\epsilon_F)$.

In summary we have studied the formation of a magnetic moment in quantum
point contacts and its relation with the observed anomalies in transport
experiments. The Hartree-Fock picture presented here
accounts for the occurrence magnetic moments when resonant states of the QPC
are aligned with the Fermi level. For a moderate local interaction, the
stable local magnetic moments are related only to the first longitudinal
resonance of each transverse mode, i.e., at the beginning of each
conductance step, in agreement with the results of Reilly \etal~%
\cite{aust1}. However, the behavior of the conductance at each conductance
step will depend of the particular details of the QPC potential. These
results are robust for QPC that generate a flat potential along the contact
axis. By mapping the problem onto a
one-channel model, we calculated the temperature and gate voltage dependence
of the conductance that are in good agreement with the experimental
observation. Our results support the idea of Kondo effect in the point
contact however our proposal differs from previous ones: in the work by Meir 
\etal~\cite{meir1} it is assumed that as the gate voltage is
increased, the magnetic moment is formed well before the conductance
increases although it is essentially decoupled from the electron gas. In our
approach the conductance increase in accompanied by the formation of the
magnetic moment and this occurs each time a new channel starts to 
participate in the transport trough the point contact. 

A large magnetic field may destroy the Kondo correlations and could be used
as an external probe to control the spin fluctuations. However, in the
experiments of  Cronenwett \etal \cite{crone} the Kondo
temperature has been estimated to be as large as 10 K and consequently only
huge magnetic field could destroy the Kondo correlations for all the gate
voltages. In a recent publication \cite{graham}, other GaAs point contacts
that for zero magnetic field present the $0.7$ anomaly, for fields of $15$
Tesla show a conductance plateau at $e^{2}/h$ with a structure that clearly
resembles a resonance at the edge of the first conductance step in agreement
with the prediction of the model presented above.

\acknowledgments
We thank B. Alascio for stimulating discussions. This work
was partially supported by the CONICET and ANPCYT, grants N. 02151 and 99
3-6343


\begin{thebibliography}{0}

\bibitem{wees1988}  
\Name{ B.J. van Wees, H. van Houten, C.W.J. Beenakker, J.G. Williamson, L.P. Kouwenhoven, D. van der Marel, \and C.T. Foxon}
\REVIEW{Phys. Rev. Lett.}{60}{1988}{848}.
  

\bibitem{cuant2}  
\Name{D.A. Wharam, \etal}
\REVIEW{J. Phys. C}{21}{1988}{L209}.

\bibitem{data}  
  \Name{S. Datta}
  \Book{Electronic Transport in Mesoscopic Systems}
  \Publ{Cambridge University, Cambridge}
  \Year{1995}.

\bibitem{aust2}
\Name{D.J. Reilly, G.R. Facer, A.S. Dzurak, B.E. Kane, R.G. Clark, P.J. Stiles, R.G. Clark, A.R. Hamilton, J.L. O'Brien, N.E. Lumpkin, L.N. Pfeiffer, \and K.W. West}
 \REVIEW{Phys. Rev. B}{63}{2001}{R121311}.


\bibitem{cerosiete}  
\Name{K.J. Thomas, J.T. Nicholls, M.Y. Simmons, M. Pepper, D.R. Mace, \and D.A. Ritchie}
\REVIEW{Phys. Rev. Lett.}{77}{1996}{135}.

\bibitem{cerosiete1} 
  \Name{K.J. Thomas, J.T. Nicholls, N, J. Appleyard, M.Y. Simmons, M. Pepper, D.R. Mace, W.R. Tribe, \and D.A. Ritchie}
  \REVIEW{Phys. Rev. B}{58}{1998}{4846}.


\bibitem{crone}  
  \Name{S.M. Cronenwett, H.J. Lynch, D. Goldhaber-Gordon, L.P. Kouwenhoven, C.M. Marcus, K. Hirose, N.S. Wingreen, \and V. Umansky}
  \REVIEW{Phys. Rev. Lett.}{88}{2002}{226805}.


\bibitem{aust1}  
  \Name{D.J. Reilly, T.M. Buehler, J.L. O'Brien, A.R. Hamilton, A.S. Dzurak, R.G. Clark, B.E. Kane, L.N. Pfeiffer, \and K.W. West}
  \REVIEW{Phys. Rev. Lett.}{89}{2002}{246801}.


\bibitem{Hewson}   
  \Name{A.C. Hewson}
  \Book{The Kondo problem to heavy fermions}
  \Publ{Cambridge University, Cambridge} 
  \Year{1993}.

\bibitem{Berggren} 
 \Name{K.-F. Berggren \and I. I. Yakimenko}
 \REVIEW{Phys. Rev. B}{66}{2002}{085323}.

\bibitem{meir2}  
  \Name{K. Hirose, Y. Meir, \and N.S. Wingreen}
  \REVIEW{Phys. Rev. Lett.}{90}{2003}{026804}.

\bibitem{meir1}  
  \Name{Y. Meir, K. Hirose, \and N.S. Wingreen}
  \REVIEW{Phys. Rev. Lett.}{89}{2002}{196802}.

\bibitem{Cornaglia2003} 
  \Name{P.S. Cornaglia,  C.A. Balseiro, \and M. Avignon}
  \Publ{unpublished}.

\bibitem{Anderson}  
  \Name{P.W. Anderson}
  \REVIEW{Phys. Rev.}{124}{1961}{41}.

\bibitem{pasw}  
  \Name{H.M. Pastawski}
  \REVIEW{Phys. Rev. B}{44}{1991}{6329};  
  \SAME{46}{1992}{4053}.


\bibitem{NRG}  
  \Name{K.G. Wilson}
  \REVIEW{Rev. Mod. Phys.}{47}{1975}{773}.

\bibitem{NRG1}
  \Name{H.R. Krishna-murthy, J.W. Wilkins, \and K.G. Wilson}
  \REVIEW{Phys. Rev. B}{21}{1980}{1044}.

\bibitem{nos}  
  \Name{P.S. Cornaglia \and C.A. Balseiro}
  \REVIEW{Phys. Rev. Lett.}{90}{2003}{216801}.

\bibitem{graham}
  \Name{A.C. Graham, K.J. Thomas, M. Pepper, N.R. Cooper, M.Y. Simmons \and D.A. Ritchie}
  \REVIEW{Phys. Rev. Lett.}{91}{2003}{136404}.
  
\end{thebibliography}
\end{document}